\documentclass[
reprint,
superscriptaddress, 
amsmath,amssymb,
aps,
prb,
prstab,
floatfix,
longbibliography
]{revtex4-2}

\usepackage[caption=false]{subfig}
\usepackage{multirow}
\usepackage{graphicx}
\usepackage{dcolumn}
\usepackage{bm}
\usepackage{threeparttable}
\usepackage{color}
\usepackage[dvipsnames]{xcolor}
\usepackage{chemformula}
\usepackage{xspace}
\usepackage{orcidlink}
\usepackage{amsmath}
\usepackage{tabularx}
\usepackage[normalem]{ulem}

\usepackage{amssymb}
\usepackage[capitalize]{cleveref}
\usepackage{booktabs}

\usepackage{pdfpages} 
\usepackage{pgffor}   

\makeatletter
\AtBeginDocument{\let\LS@rot\@undefined}
\makeatother

\def\supplementfilename{Suppl.pdf}

\pdfximage{\supplementfilename}
\def\numbersupplementpages{\the\pdflastximagepages}

\newcommand{\rbmnf}{\ch{RbMnF3}\xspace}
\newcommand{\mnf}{\ch{MnF2}\xspace}
\newcommand{\fef}{\ch{FeF2}\xspace}
\newcommand{\knif}{\ch{KNiF3}\xspace}

\newcommand*{\rom}[1]{\expandafter\@slowromancap\romannumeral #1@}

\definecolor{AMK}{RGB}{220, 10, 10}
\definecolor{new}{RGB}{0, 0, 0}
\definecolor{old}{RGB}{150, 150, 150}
\definecolor{AEF}{RGB}{181, 58, 212}

\bibpunct{[}{]}{;}{n}{}{}                                                                                                                                                                                                       

\hypersetup{}

\begin{document}
\preprint{APS/123-QED}

\title{Role of magnon-magnon interaction in optical excitation\\of coherent two-magnon modes}

\author{E.~A.~Arkhipova\,\orcidlink{0009-0002-7939-9749}}
    \email{elizaveta.arhip@mail.ioffe.ru}
    \affiliation{Ioffe Institute, 194021 St.\,Petersburg, Russia}
    
\author{A.~E.~Fedianin\,\orcidlink{0000-0002-5093-6830}}
    \affiliation{Ioffe Institute, 194021 St.\,Petersburg, Russia}
    
\author{I.~A.~Eliseyev\,\orcidlink{0000-0001-9980-6191}}
    \affiliation{Ioffe Institute, 194021 St.\,Petersburg, Russia}

\author{R.~M.~Dubrovin\,\orcidlink{0000-0002-7235-7805}}
     \affiliation{Ioffe Institute, 194021 St.\,Petersburg, Russia}
     
\author{P.~P.~Syrnikov}
    \affiliation{Ioffe Institute, 194021 St.\,Petersburg, Russia}

\author{V.~Yu.~Davydov\,\orcidlink{0000-0002-5255-9530}}
    \affiliation{Ioffe Institute, 194021 St.\,Petersburg, Russia}       
    
\author{A.~M.~Kalashnikova\,\orcidlink{0000-0001-5635-6186}}
    \affiliation{Ioffe Institute, 194021 St.\,Petersburg, Russia}

\date{\today}

\begin{abstract}

Two-magnon modes are terahertz-frequency magnetic excitations in antiferromagnets, governed by exchange interactions, involving magnons from the entire Brillouin zone \textcolor{new}{and dominated by zone-edge magnons}. 
The ability to couple to light promotes two-magnon modes as contenders for ultrafast optical manipulation of the magnetic state, beyond conventional zone-center magnonics.
While magnon-magnon interactions are known to critically shape the two-magnon line in spontaneous Raman scattering spectra, their role in coherent time-domain excitations remains unexplored.  
We report a detailed experimental and theoretical study of the influence of magnon-magnon interactions on coherent two-magnon modes in a cubic antiferromagnet excited via Impulsive Stimulated Raman scattering.
We reveal the nontrivial evolution of coherent magnetic dynamics in the time domain and the corresponding spectrum, and compare it with the spontaneous Raman scattering spectrum.
By extending the spin-correlations based theory for two-magnon modes, we derive a unified description of their spectra in Raman Scattering and Impulsive Stimulated Raman Scattering and highlight the role of magnon-magnon interactions.

\end{abstract}

\maketitle


Modern technologies place increasingly high demands on data read/write speeds, storage density, and energy efficiency.
This is particularly true for the development of elements for analogue logic and neuromorphic computing~\cite{Kurenkov2020}. 
Magnonics deals with exciting and controlling spin waves (magnons) and is a highly promising field for addressing these problems~\cite{Barman2021,Flebus2024}, with the generation and control of ultra-high-frequency, short-wavelength spin waves being a key fundamental challenge.
Antiferromagnets offer a promising solution, as they exhibit terahertz-frequency magnetic excitations called two-magnon modes~(2M).
These modes are primarily determined by perturbations in the exchange interaction and correspond to the collective excitation of pairs of magnons with frequencies $\Omega_\mathbf{k}$ and opposite wave vectors $\pm\mathbf{k}$ throughout the Brillouin zone.
Unlike single-magnon modes with large $\mathbf{k}$, 2M modes have zero total momentum. 
As a result, they are Raman-active~\cite{fleury1968scattering,thorpe1970two,lockwood1975two} and can even be infrared-active~\cite{Richards1967}, effectively interacting with electromagnetic radiation, with the largest contribution coming from the zone-boundary magnon pairs.
This enables their driving in a coherent manner by short optical~\cite{Zhao2004,Bossini2016,Formisano2024} and mid-infrared laser pulses~\cite{Schönfeld2025,Shan2024}, opening up a prospective chapter in the optical manipulation of magnetic states \textcolor{new}{with several ambitious proposals of how coherent zone-edge magnon pairs could be exploited in quantum computing~\cite{Zhao2004} and magnonics~\cite{hegstad2025ultrafast,fabiani2021supermagnonic}}.

Magnon-magnon interactions \textcolor{new}{are governed by exchange coupling and} play a determining role in magnetic dynamics.
They underlie fundamental phenomena such as the softening of magnon eigenfrequencies with temperature~\cite{Oguchi1961}, angular momentum dissipation, and the associated attenuation of spin waves~\cite{Zakeri2007}, and they constitute one of the key mechanisms for establishing thermodynamic equilibrium within the spin subsystem~\cite{Cherepanov1993}.
The role of these interactions is also actively discussed in emergent fields such as femtomagnetism~\cite{Kirilyuk2010,Schönfeld2025}, room-temperature Bose-Einstein condensation of magnons~\cite{Dzyapko2017}, and cavity~\cite{Parvini2020,Parvini2025,Yuan2020} and quantum~\cite{serha2025quant} magnonics.
For 2M modes, the effect of magnon-magnon interactions is particularly pronounced, as confirmed in numerous experimental and theoretical studies~\cite{Elliott1969,canali1992theory,balucani1973theory,Powalski2015,Powalski2018,Walther2023}. 
The spontaneous Raman Scattering (RS) spectrum of a two-magnon mode cannot be adequately described without taking these interactions into account, not only in terms of the linewidth but also in terms of the distribution of spectral amplitudes.

While RS provides information on the incoherent ensemble of 2M modes, coherent 2M modes can be accessed in the time domain via Impulsive Stimulated Raman Scattering~(ISRS), yielding not only frequency and amplitude but also phase information.
\textcolor{new}{In} several experimental studies for \mnf and \fef~\cite{Zhao2004,Zhao2006} and for \knif~\cite{Bossini2016}, an apparent difference in 2M spectra in RS and ISRS experiments \textcolor{new}{was noted}.
In Ref.~\cite{Fedianin2024}, a comparative theoretical analysis of RS and ISRS confirmed this difference for \rbmnf, without accounting for magnon-magnon interactions.
Since the influence of the magnon-magnon interaction is clearly manifested in the RS spectra, its impact on an ensemble of coherent 2M modes driven by ISRS and on the observed dynamics is anticipated but remains unexplored.

In this Letter, we report on a detailed comparative experimental study of two magnon modes observed in RS and ISRS in a model antiferromagnet \rbmnf. 
Supported by a semi-quantum theory that describes the coupling of 2M modes to light in the presence of magnon-magnon interactions, our study reveals their pronounced impact on coherent 2M modes in an ISRS experiment, leading to specific spectral features.
We successfully describe the experimental RS and ISRS spectra and confirm their distinct features. 
We also show that the difference persists over a broad temperature range below the ordering temperature.

 \begin{figure}[!t]
     \centering
     \includegraphics[width=1\columnwidth]{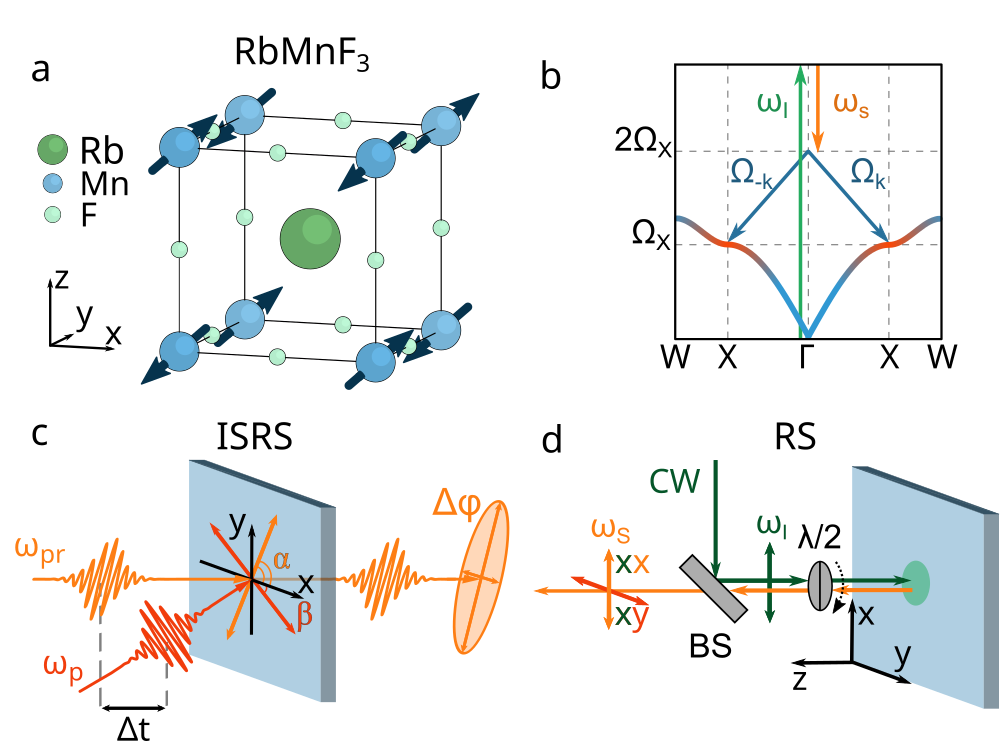}
     \caption{
         \label{fig:experimental_setup}
         (a) Crystal and magnetic structure of \rbmnf. Blue and red arrows indicate spins in two sublattices aligned along the $\langle111\rangle$ direction.\textcolor{new}{The figure was created using VESTA~\cite{momma_vesta3_threedimensional_2011}.}
         (b) Schematic representation of magnon dispersion and light interaction with a two-magnon mode.
         The figure depicts inelastic light scattering via a two-magnon excitation at the X point of the Brillouin zone. 
         The color gradient schematically represents an effect of magnon-magnon interaction.
         (c) The time-resolved pump-probe experimental scheme: 
         linearly polarized at an angle $\beta$ optical pump pulse excites spin dynamics via ISRS, which is detected via transient ellipticity of the optical probe pulse. 
         $\Delta t$ is time delay between pump and probe. 
         (d) RS experimental scheme in back-scattering geometry with $z \parallel [001]$.
         Incident continuous-wave (CW) beam is polarized along the $x \parallel [100]$ axis. 
         In the parallel ($xx$) and cross-polarized ($xy$) configurations of the scattered beam is polarized along the $x$ and $y \parallel [010]$ axes, respectively.
    }
 \end{figure}

\rbmnf is a cubic antiferromagnet [Fig.~\ref{fig:experimental_setup}(a)] with a perovskite-type structure (space group~$Pm\overline{3}m$) and a lattice constant $a = 4.234$\,\AA~\cite{Windsor1966}.
This is a Heisenberg antiferromagnet with spin $S = 5/2$, an exchange constant $J = 0.588$\,meV, a N{\'e}el temperature $T_{\mathrm{N}}=83$\,K~\cite{fleury1968scattering}, and very weak magnetic anisotropy~\cite{LpezOrtiz2014}. 
The absence of Raman-active phonons enables the observation of 2M modes not obscured by superposition with intense phonon contributions in both RS~\cite{fleury1968scattering} and ISRS experiments~\cite{Formisano2024}.
This material is an insulator with weak optical absorption that starts at approximately 2.4\,eV~\cite{Mehra1967}.
The \rbmnf sample with 800~$\mu$m thickness was cut and polished in the $(001)$ plane from a single crystal grown by the Czochralski method.

In both RS and ISRS experiments, the 2M modes are coupled to light via an elementary inelastic scattering process, schematically shown in Fig.~\ref{fig:experimental_setup}(b).
This process is mediated by a light-induced perturbation of the exchange coupling.
In RS, the incoherent population of thermally excited 2M modes is measured.
The spectra result from the time-integrated detection of multiple scattering events with random phases, appearing as a frequency-shift relative to the incident monochromatic light.
In ISRS, the coherent 2M modes are driven by a short laser pump pulse and are monitored in the time domain through a transient polarization state of a short probe pulse after a delay time.
The dynamics of the two-magnon mode is associated with a laser-induced perturbation of the exchange interaction~\cite{mentink2015ultrafast}, and manifests itself in a time-dependent change in the dielectric permittivity and induced optical anisotropy~\cite{Fedianin2023}. 

To study coherent 2M modes excited via ISRS, we used the two-colour pump-probe technique in the transmission geometry [Fig.~\ref{fig:experimental_setup}(c)].
The pump-probe experiments were carried out using a Ti:Sapphire laser system (REUS, Avesta) with a central photon energy of 1.55\,eV, a pulse duration of 35\,fs, and a repetition rate of 1\,kHz. 
The pulse from the laser was split into probe ($\hbar \omega_{\mathrm{pr}} = 1.55$\,eV) and pump beams.
To tune the pump pulse to an energy $\hbar \omega_{\mathrm{p}} = 1.19$\,eV, which is far below the absorption edge and corresponds to negligible absorption in \rbmnf, we used an optical parametric amplifier. 

The sample was placed in the continuous flow liquid helium cryostat. 
Linearly polarized pulses were used, the polarization angles $\alpha$ and $\beta$ for the probe and pump pulses, respectively, relative to the $x \parallel [100]$ axis of \rbmnf [see Fig.~\ref{fig:experimental_setup}(c)]. 
The pump spot diameter at the sample was about 55\,$\mu$m.
The pump pulse fluence was varied from $F_{p} = 60$\,mJ/cm$^2$ to 160\,mJ/cm$^2$, and an incidence angle was of about 10$^\circ$. 
Probe pulses with a fixed fluence of 0.13\,mJ/cm$^2$ were focused on the sample into a spot diameter of 20\,$\mu$m at normal incidence.

\textcolor{new}{To observe the ISRS effect, the pump and probe pulse duration must be shorter than the oscillation period of the studied modes~\cite{Misochko2016,Imasaka2018,Formisano2024}.}
The intensity full width at half maximum for the pump (90\,fs) and the probe (50\,fs) right in front of the sample were measured using an autocorrelator. 
\textcolor{new}{In the experiments, the pump and probe pulses were both close to Gaussian, with} actual durations of $\tau_\mathrm{p} = 64$\,fs and $\tau_\mathrm{pr} = 32$\,fs.
Coherent spin dynamics was detected through the change in the pump-induced ellipticity $\Delta\phi$ of the probe pulse as a function of the time delay $\Delta t$ between the pump and probe pulses, using a balanced detection scheme.
To ensure sufficient resolution, the time-delay step was 27\,fs, and the scanning was performed over 12\,ps~\cite{supp_mat}.
The corresponding spectral resolution was 2.7\,cm$^{-1}$. 

For comparison of the two-magnon modes excited in the coherent regime to the noncoherent ones, RS measurements were performed using a Horiba LabRAM HR Evolution UV-VIS-NIR (Horiba) spectrometer equipped with a confocal microscope. 
The spectra were measured using continuous-wave excitation with a 2.33\,eV line of a Nd:YAG laser (Laser Quantum), and the power of the incident light on the sample was 12\,mW.
Spectra were recorded in backscattering geometry [Fig.~\ref{fig:experimental_setup}(c)] using a 2400\,lines/mm grating, providing a spectral resolution of 0.3\,cm$^{-1}$ and a nitrogen-cooled charge-coupled device detector, while an Olympus LMPFLN 50× (NA = 0.5) long working-distance objective lens was used to focus the incident beam into a spot with a diameter of 2\,µm.
The required polarization component was selected using a Glan-Taylor prism. 
In angle-resolved polarized Raman measurements, the polarization of the incident and scattered beams was varied by rotation of an achromatic halfwave plate.

\begin{figure}[!t]
     \centering
     \includegraphics[width=1\columnwidth]{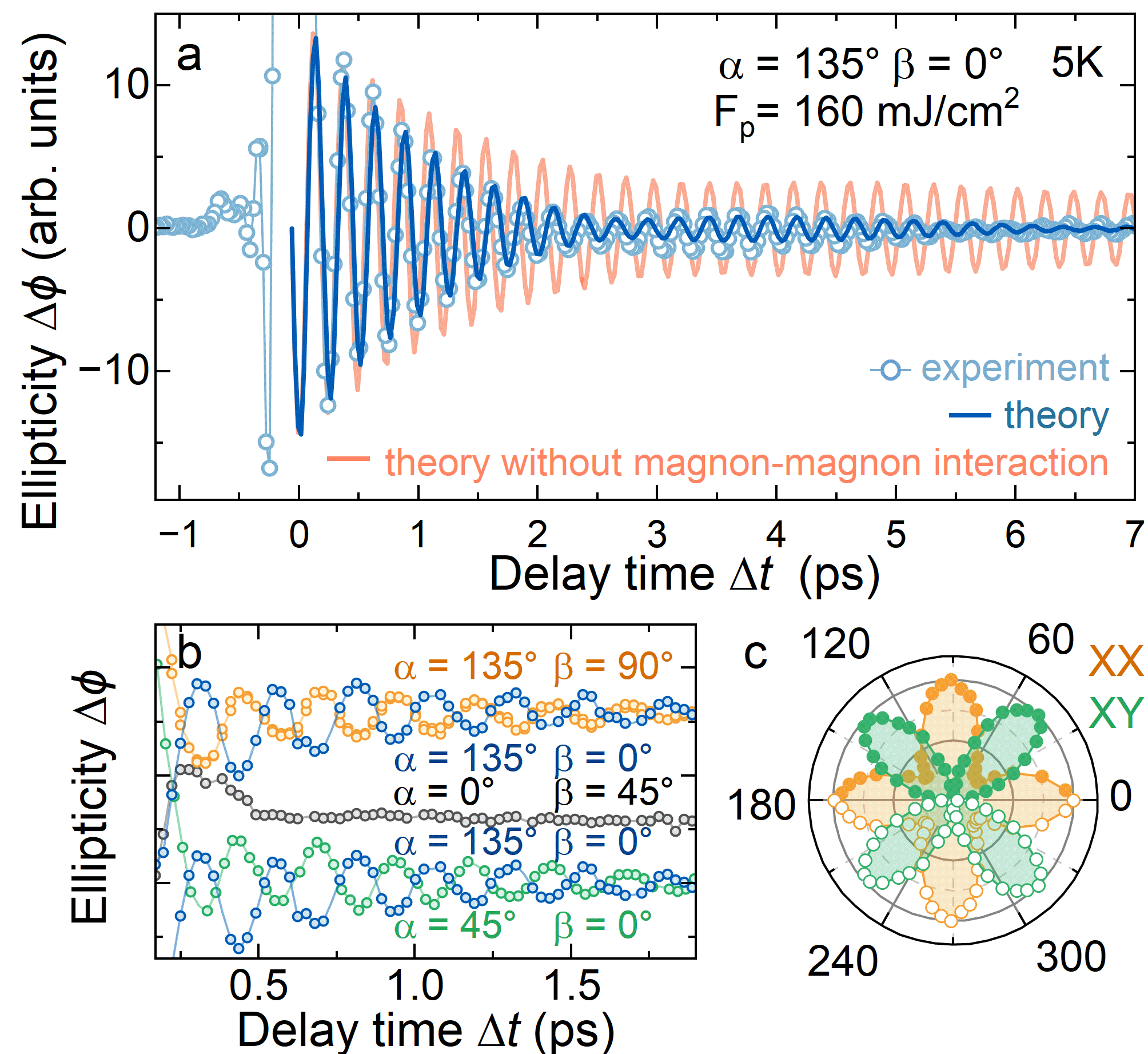}
     \caption{
         \label{fig:Time_trace}
          (a)~Laser-induced probe ellipticity $\Delta\phi$ as a function of the delay time $\Delta t$ (open blue symbols) measured in \rbmnf at $T = 5$\,K. 
          Lines -- The inverse Fourier transform of the calculated ISRS spectra of 2M mode [Eq.~\eqref{eq:ellipticity}] obtained taking into account the magnon-magnon interaction (blue line), and without magnon-magnon interaction (red line).          
          (b)~Transient probe ellipticity $\Delta\phi(\Delta t)$ measured at different combinations of the pump $\beta$ and probe $\alpha$ polarization angles.
          (c)~Polarization dependences of the intensity of the 2M mode line in measured RS spectra in parallel [$z(xx)\overline{z}$] (solid orange symbols) and crossed [$z(xy)\overline{z}$] (solid green symbols) configurations. 
          The open symbols are the experimental data reflected relative to the horizontal axis, assuming the angular symmetry. 
    }
 \end{figure}

\begin{figure}[!t] 
     \centering
     \includegraphics[width=\columnwidth]{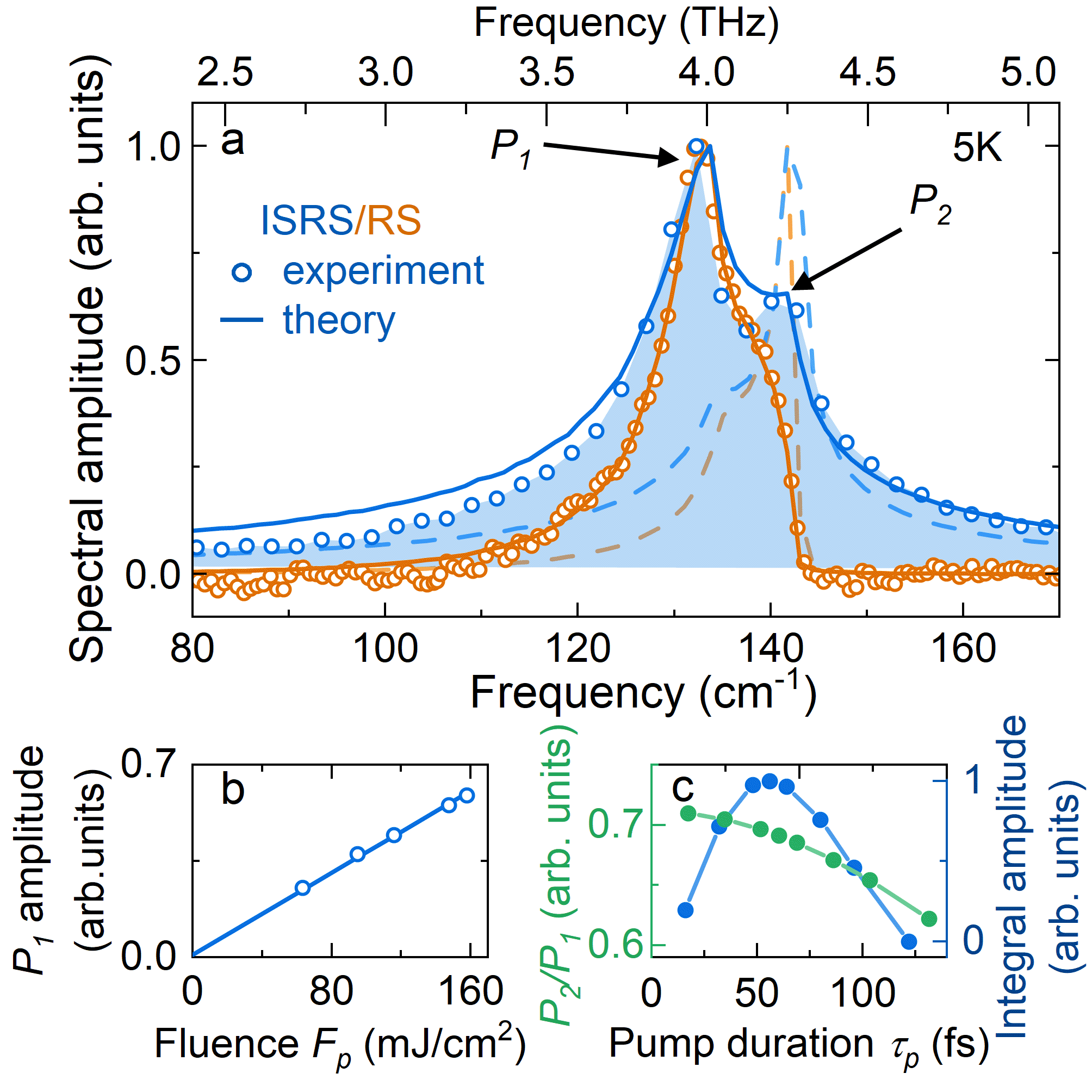}
     \caption{
         \label{fig:2MM_spectra}
          (a)~2M mode spectra at 5\,K. 
          Open blue symbols are the Fourier transform of experimental transient probe ellipticity [Fig.~\ref{fig:Time_trace}(a)], solid blue line is the calculated ISRS 2M spectrum taking into account the magnon-magnon interaction [Eq.~\eqref{eq:ellipticity}]. 
          Open orange symbols are the experimental RS 2M spectrum, solid orange line is the calculated RS spectrum taking into account the magnon-magnon interaction [Eq.~\eqref{eq:scattering}].
          Dashed lines are a calculated spectrum of 2M mode in RS (orange) and ISRS (blue) without magnon-magnon interaction. 
          All spectra were normalized by the maximum of magnitude. 
          (b)~Pump fluence $F_{p}$ dependence of the FFT ISRS amplitude. 
          The line is a linear fit. 
          (c)~Integral amplitude (blue symbols) of the calculated ISRS 2M spectrum [Eq.~\eqref{eq:ellipticity}] and ratio of the amplitudes of the features $P_{2}/P_{1}$ (green symbols) vs pump duration $\tau_\mathrm{p}$.
          The lines are guides to the eye.         
     }
 \end{figure}

Figure~\ref{fig:Time_trace}(a) shows the representative time trace obtained in the ISRS experiment at $T = 5$\,K.
The transient ellipticity oscillates with the main frequency of 4\,THz, which is close to the frequency of the 2M mode at low temperature~\cite{fleury1968scattering}. 
The dependence of the signal on the polarization of the pump and probe pulses, shown in Fig.~\ref{fig:Time_trace}(b), fully agrees with the ISRS selection rules for a cubic antiferromagnet, which state that the pump and probe pulses should be polarized along and at 45$^{\circ}$ to one of the main crystallographic axes, respectively, for the most efficient excitation and detection~\cite{Formisano2024, Fedianin2023}. 
Therefore, we conclude that the observed transient probe polarization dynamics stems from the coherent 2M mode excited by the pump pulse.  
It is important to note that the transient ellipticity exhibits well-resolved beatings, which we attribute to the fact that the 2M modes from various points in the Brillouin zone contribute to the signal.
The \textcolor{new}{Fast} Fourier Transform \textcolor{new}{(FFT)} for the transient ellipticity signal is shown in Fig.~\ref{fig:2MM_spectra}(a) (open blue symbols).
The beatings in the time domain manifests itself as two spectral features in the frequency domain, which we label as $P_{1} = 132.5$\,cm$^{-1}$ (3.97\,THz) and $P_{2} = 141.25$\,cm$^{-1}$ (4.23\,THz).
We observed that the amplitude of the transient ellipticity oscillations grows linearly with pump fluence, in agreement with the ISRS process [Fig.~\ref{fig:2MM_spectra}(b)]. 
The spectrum, in turn, does not change with the pump fluence, confirming the absence of additional heating.
\textcolor{new}{We note that the choice of a specific Fourier transform algorithm does not affect the results.
The features of the two-magnon mode in the spectra obtained using the FFT demonstrate high stability.}

The 2M mode RS spectrum measured in $z(xx)\overline{z}$ configuration at $T = 5$\,K is shown in Fig.~\ref{fig:2MM_spectra}(a) (open orange symbols).
Azimuthal dependences of the normalized RS intensity in the parallel ($xx$) and crossed-polarized ($xy$) configurations [Fig.~\ref{fig:Time_trace}(c)] are
consistent with the RS selection rules for a cubic antiferromagnet~\cite{Elliott1969}. 
The spectrum is in very good agreement with those reported earlier for \rbmnf, and its shape is governed by the magnon density of states and magnon-magnon interactions~\cite{fleury1968scattering}.
However, the ISRS spectrum does not follow the RS one, and the difference extends well beyond a simple broadening that could be ascribed to the different experimental resolutions of the two techniques.

The earlier developed~\cite{Fedianin2024} single-framework theory of the RS and ISRS from 2M modes, which neglected the magnon-magnon interaction, yield the spectra shown by the dashed lines in Fig.~\ref{fig:2MM_spectra}(a), and clearly do not provide a proper description of either of the experimental spectra, despite capturing a broadening of the ISRS one.
The discrepancy between these theoretical and experimental results directly demonstrates the role of magnon-magnon interactions in both regimes. 
The theory that neglects magnon-magnon interactions also fails to reproduce the experimental time-domain dynamics, as shown by the light red line in Fig.~\ref{fig:Time_trace}(a). 
The oscillations exhibit identical initial phases, but a gradual divergence occurs on a picosecond timescale, which correlates with the discrepancy in a frequency domain.
Therefore, to correctly describe the experimental results, it is necessary to incorporate into the unified theory for RS and ISRS~\cite{Fedianin2024} the approach that takes into account the magnon-magnon interaction developed for RS~\cite{davies1971spin}.

 \begin{figure}[!t]
     \centering
     \includegraphics[width=1\columnwidth]{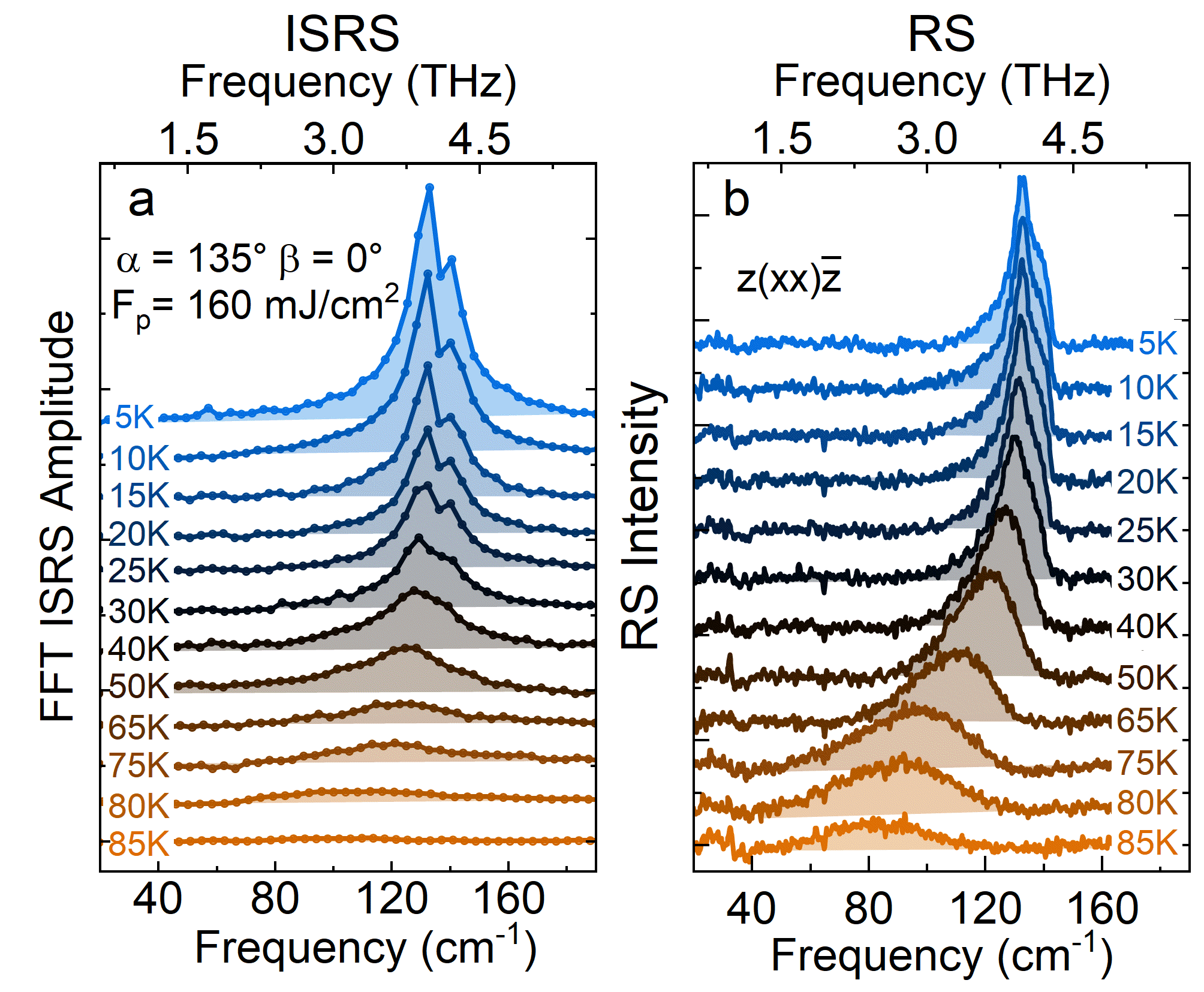}
     \caption{
         \label{fig:TempDep}
         Experimemntal 2M mode spectra as obtained at different temperatures from
         (a)~the Fourier transform of experimental transient probe ellipticity, and
         (b)~RS measurements in $z(xx)\overline{z}$ configuration.
    }
 \end{figure}

The spin subsystem of \rbmnf is described by the Heisenberg model $\hat{H}=J \sum_{i,\delta} \hat{\mathbf{S}}_{i} \cdot \hat{\mathbf{S}}_{i+\delta}$, where
$\hat{\mathbf{S}}_{i}=\hat{\mathbf{S}}(\mathbf{R}_i)$ is a spin operator at a position $\mathbf{R}_i$ in the lattice, the subscript $\delta$ denotes a vector $\boldsymbol{\delta}=(\delta^x,\delta^y,\delta^z)^T$ to the nearest neighbor.
In the linear spin wave approach, the spin product can be replaced with the spin correlation pseudovector $\mathbf{\hat{K}_k}$, which diagonalizes the Hamiltonian~\cite{Formisano2024}.
Considering the forth order of the spin operators product $\hat{S}^{-}_{i}\hat{S}^{+}_{i} \hat{S}^{+}_{i+\delta}\hat{S}^{-}_{i+\delta}$ achieved after bosonization, we derive the two-magnon Hamiltonian with magnon-magnon interaction (Sec. II of Suppl. Mater.~\cite{supp_mat})
\begin{align}
    \hat{H}&=\hbar \sum_{\mathbf{k}} 2\Omega_{\mathbf{k}} \hat{K}^z_{\mathbf{k}}+\hat{H}_1,\label{eq:Hamiltonian}\\
    \hat{H}_1 &= -I\sum_{\mathbf{k},\mathbf{q}} \gamma_{\mathbf{k}-\mathbf{q}} \left(1+\frac{1}{\varepsilon_\mathbf{k}\varepsilon_\mathbf{q}}\right) \hat{K}^{+}_{\mathbf{k}} \hat{K}^{-}_{\mathbf{q}},\label{eq:MM_Hamiltonian}\\
    2\Omega_{\mathbf{k}}&\equiv2JSn_a\varepsilon_{\mathbf{k}},\quad I\equiv\frac{J n_a}{2N},\quad\gamma_{\mathbf{k}}\equiv\frac{1}{n_a}\sum_{\boldsymbol{\delta}} e^{-i\mathbf{k}\cdot\boldsymbol{\delta}},\nonumber
\end{align}
where $n_a$ is the number of the nearest neighbors, $N$ is the number of wavevectors, and $\varepsilon_\mathbf{k}\equiv\sqrt{1-\gamma_{\mathbf{k}}^2}$ is the normalized magnon energy.
While the non-interacting part of the Hamiltonian represents the ensemble of oscillators, $\hat{H}_1$ bounds them with each other.
\textcolor{new}{Note that $\hat{H}_1$ accounts for exchange coupling as a mechanism of magnon-magnon interactions.
To keep the problem solvable analytically, the magnon dispersion is considered without accounting for the magnon-magnon interactions.
We keep only the components of the laser-induced exchange interaction perturbation that are linear in $\mathbf{K_k}$, as the corresponding changes in the magnon-magnon interaction are the next order of magnitude (see Eq. (8) in Sec. III of Suppl. Mater.~\cite{supp_mat}).}

As shown in Ref.~\cite{Fedianin2024}, the RS cross-section is found through the imaginary part of the Green function of $\mathbf{\hat{K}_k}$, while the spectrum of the transient ellipticity in ISRS is given by its absolute value. 
However, in the presence of the magnon-magnon interaction, the expectation value of the RS cross-section cannot be calculated using a trace of the Green function.
Therefore, we have to sum the Green function $G(\hat{K}^{-}_{\mathbf{k}},\hat{K}^{+}_{\mathbf{q}}|\omega)$ over all pairs of wave vectors $\mathbf{k}$ and $\mathbf{q}$, as detailed in Sec. III Suppl. Mater.~\cite{supp_mat}.
Considering [$z(xx)\overline{z}$] RS configuration, the expression for the scattering intensity takes the form derived earlier in~\cite{davies1971spin,balucani1973theory,canali1992theory}
\begin{align}
    \sigma(\omega)\!&\propto\!\mathrm{Im}\!\left\{\!\frac{L_2\!+\!I\left(L_0 L_2\!-\!L_1 L_1\right)}{1\!+\!I\left(L_0\!+\!L_2\right)\!+\!I^2\left(L_0 L_2\!-\!L_1 L_1\right)}\!\right\}\!,\label{eq:scattering}\\
    L_m(\omega)&\equiv\sum_{\mathbf{k}} \eta^{xxX}_{\mathbf{k}} \eta^{xxX}_{\mathbf{k}} \varepsilon_{\mathbf{k}}^{2-m} G^{(0)}_{\mathbf{kk}}(\omega),\nonumber
\end{align}
where $\eta^{\nu\nu X}_\mathbf{k}$ has the meaning of the normalized component of the Raman tensor, with $\nu$ indices in real space denoting the incident and scattered light polarizations and $X$ refers to the $\mathbf{\hat{K}}$ component in hyperbolic two-magnon space, $G^{(0)}_{\mathbf{kk}}(\omega)$ denotes $G^{(0)}(\hat{K}^{-}_{\mathbf{k}},\hat{K}^{+}_{\mathbf{k}}|\omega)$, which is the Green function for the non-interacting case.

\textcolor{new}{
In the case of ISRS, the strong electric field of the pump pulse provides an additional contribution to the exchange energy $J$ during the pulse~\cite{mentink2015ultrafast}, which brings the system out of equilibrium and leads to the emergence of dynamics.
Therefore,} the finite duration of the pump pulse affect the efficiency of the excitation of two-magnon modes.
Moreover, modulation of the probe ellipticity with the largest amplitude arises at $45^\circ$ degree polarization, leading to a different combination of the Raman tensor components entering the resulting expression. 
Considering the experimental ISRS geometry with the pump and probe polarization angles $\beta=0$ and $\alpha=135^\circ$, the expression for the spectrum of the transient ellipticity takes a form
\begin{align}
    \Delta\phi(\omega)&\propto \left|\frac{\tilde{L}_2+I\left(L_0\tilde{L}_2-L_1\tilde{L}_1\right)}{1+I\left(L_0+L_2\right)+I^2\left(L_0 L_2-L_1 L_1\right)}\right|,\label{eq:ellipticity}\\
    \tilde{L}_{m}(\omega)&\equiv\sum_{\mathbf{k}} f_{\mathbf{k}} \left(\eta^{xxX}_{\mathbf{k}}-\eta^{yyX}_{\mathbf{k}}\right) \eta^{xxX}_{\mathbf{k}} \varepsilon_{\mathbf{k}}^{2-m} G^{(0)}_{\mathbf{kk}}(\omega),\nonumber
\end{align}
where $f_{\mathbf{k}}=\tau_p \exp{\left(-\Omega_{\mathbf{k}}^2\tau_p^2\right)}$ is a factor characterizing the efficiency of two-magnon mode excitation by a Gaussian pump pulse.
Note that Eqs.~\eqref{eq:scattering} and~\eqref{eq:ellipticity} transform into the expressions for RS and ISRS spectra without magnon-magnon interactions~\cite{Fedianin2024} by substituting $I = 0$.

As shown by the solid lines in Fig.~\ref{fig:2MM_spectra}(a), Eqs.~\eqref{eq:scattering} and~\eqref{eq:ellipticity} describe both RS and ISRS spectra with good accuracy, without any fitting parameters.
For ISRS, this theory successfully reproduces the experimental time-domain dynamics, as shown by the solid blue line in Fig.~\ref{fig:2MM_spectra}(a).
The introduction of magnon-magnon interaction bounds two-magnon oscillators, and light excites the entire ensemble of oscillators at each act of scattering.
This leads to a redistribution of spectral intensity in both cases of RS and ISRS [Fig.~\ref{fig:2MM_spectra}(a)].
At the same time, polarization dependencies persist. 
Accordingly, the main reasons for the differences between the RS and ISRS spectra are those identified earlier~\cite{Fedianin2024} for the case of non-interacting 2M modes, \textcolor{new}{and we elaborate on them below.}

That is, we attribute the broadening of the ISRS spectrum and the blue-shift in the $P_{2}$ feature to several factors.
Firstly, RS probes a continuum of incoherent two-magnon modes produced by multiple scattering processes.
Secondly, the finite duration of the pump pulse affects the efficiency of the excitation of different 2M modes.
In the case of impulsive excitation, coherent dynamics is measured, which is influenced by both the amplitude and phase of the modes.
We note that previously an $\approx1$~cm$^{-1}$ shift of the 2M line in the femtosecond stimulated RS was reported in another Heisenberg antiferromagnet \knif~\cite{Batignani2015} under nonresonant excitation, and attributed to the impulsive change of exchange coupling during the pump pulse action.
In our study, the ISRS spectrum is obtained from the transient dynamics \textcolor{new}{that is measured after the pump pulse has passed,} and is governed by the equilibrium $J$ \textcolor{new}{in line with the conclusions in Ref.~\cite{Batignani2015}}.

To demonstrate the importance of the pump duration, we theoretically analyzed the dependence of the ISRS integral amplitude and the $P_{2}/P_{1}$ features magnitude ratio on the pump pulse duration.
The optimal pulse duration for excitation of a particular 2M mode is close to a fifth of its oscillation period~\cite{Formisano2024}.
For example, for $2\Omega_{k} = 4$\,THz, the optimal pulse duration will be 54\,fs, which is close to the experimental value of $\tau_\mathrm{p}$. 
Indeed, as shown in Fig.~\ref{fig:2MM_spectra}(c), the integral amplitude increases the closer the pump pulse duration is to the optimal value.
However, the ratio of spectral amplitudes of the two features $P_{2}/P_{1}$ increases with decreasing $\tau_\mathrm{p}$.
This occurs because the $P_{1}$ and $P_{2}$ features have slightly different optimal pulse durations.

Having established experimentally and confirmed theoretically the difference between RS and ISRS spectra at low temperature, we turned to a comparative study of their evolution with temperature.
Time-resolved ellipticity measurements were performed as a function of temperature~\cite{supp_mat}, and the FFT of the obtained results are plotted in Fig.~\ref{fig:TempDep}(a). 
Figure~\ref{fig:TempDep}(b) shows the RS spectra obtained at the same temperatures.
In the ISRS spectrum, the 2M mode spectral line broadens with temperature, and the high-frequency feature $P_{2}$ becomes less pronounced.
This qualitatively agrees with the behavior of the 2M line in RS spectra.
However, as the 2M line in the ISRS spectra is intrinsically broader and differs in shape compared to the line in the RS spectrum, the temperature-induced red shift of the former appears to be less evident.
In the case of ISRS, when the temperature increases above 65\,K, it is difficult to draw conclusions about the lineshape in the spectral domain, as the ellipticity signal in the time domain exhibits too few oscillations for a reliable Fourier transform~\cite{supp_mat}.

In conclusion, we reveal experimentally that magnon-magnon interactions strongly affect coherent two-magnon modes excited via ISRS in a cubic antiferromagnet \rbmnf. 
By extending the theory based on the formalism of spin-correlation pseudovector to include magnon-magnon interactions, we successfully describe coherent two-magnon spectra and time traces in the ISRS experiment.
Comparison with RS reveals that magnon-magnon interactions similarly redistribute spectral amplitude to lower energies in both cases.
The broadening of ISRS spectrum stems from the fact that the former is affected by the amplitudes and phases of coherent modes, while the phases are inaccessible in the RS experiments.
Calculations show that the pump pulse duration impacts the coherent 2M ISRS spectrum by modifying its overall magnitude and the relation between different spectral features.
Finally, we show that the revealed differences between the 2M lines in ISRS and RS spectra persist over a wide temperature range below N{\'e}el temperature and should be carefully accounted for when comparing the characteristics of the 2M modes in the two experiments.
\textcolor{new}{Our experimental results are successfully described by the theoretical model, where the laser pulse excites two-magnon modes and does not affect magnon-magnon interactions. 
Therefore, the laser pulse most efficiently excites two-magnon modes at the Brillouin zone boundary, where the density of states is maximal. 
The role of magnon-magnon interactions is in a redistribution of the spectrum of the excited modes. 
The formation of such a collective response, which includes contributions from various modes, ultimately governs the observed spin dynamics.}
This, for instance, has to be taken into consideration when designing experiments on the coherent control of 2M modes.

\section*{Supplementary Material}

See the supplementary material for the experimental transient signals used to obtain the ISRS spectra, along with a detailed derivation of the Raman scattering cross-section and the ISRS transient ellipticity that accounts for magnon-magnon interaction.

\section*{Acknowledgements}

We thank R.\,V.~Pisarev for stimulating discussions, as well as L.\,A.~Shelukhin and M.\,A.~Prosnikov for their help with experiments.
The work of E.\,A.\,A. and A.\,M.\,K. was supported by RSF grant No. 23-12-00251.








\bibliography{main}

\clearpage
\includepdf[pages={1}]{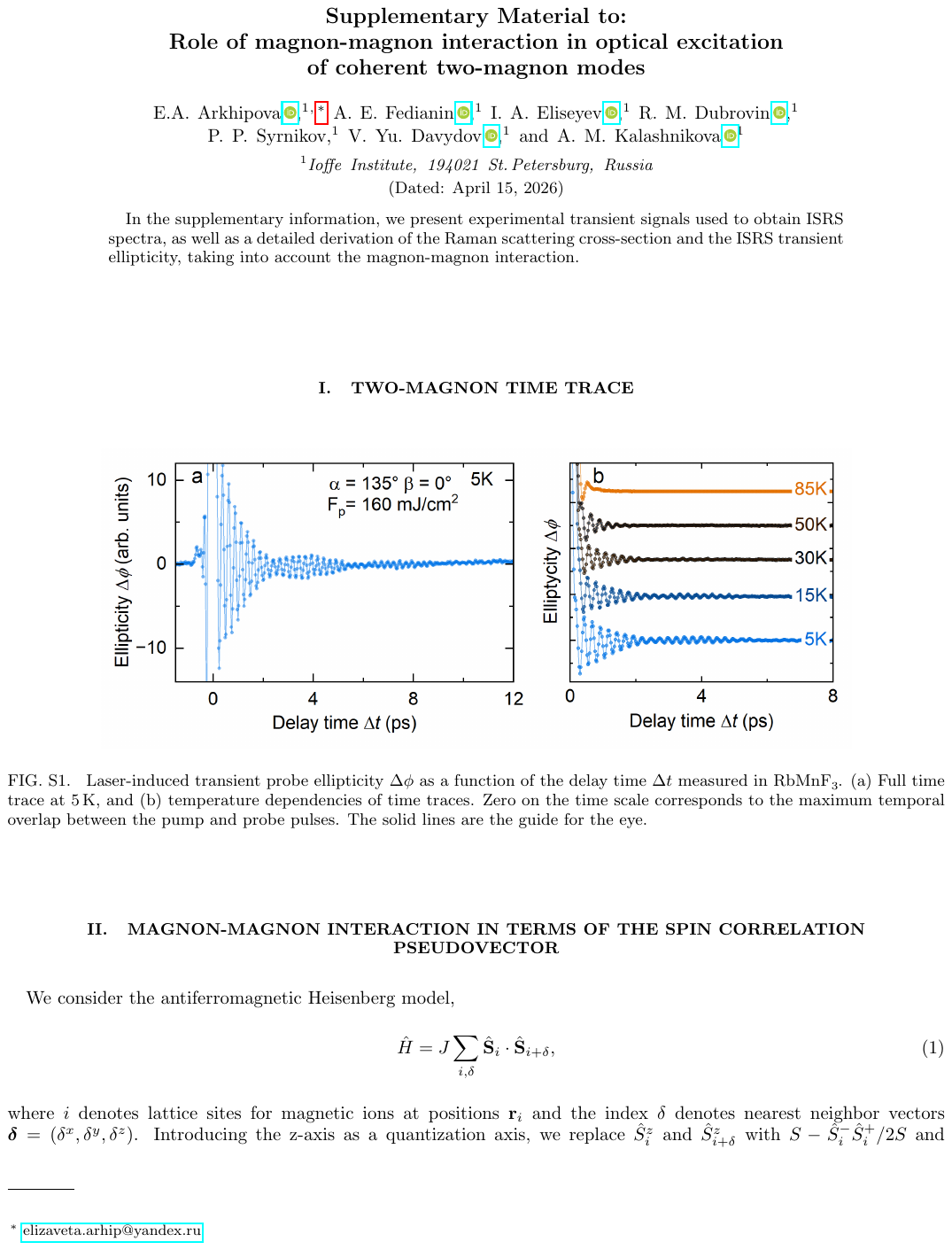}
\clearpage
\includepdf[pages={2}]{Suppl.pdf}
\clearpage
\includepdf[pages={3}]{Suppl.pdf}
\clearpage
\includepdf[pages={4}]{Suppl.pdf}
\clearpage
\includepdf[pages={5}]{Suppl.pdf}

\end{document}